\documentclass[preprints,article,accept,pdftex,moreauthors]{Definitions/mdpi} 
\usepackage{aas_macros}
\usepackage{mathtools}
\usepackage{amssymb,amsmath}
\usepackage{graphbox}
\usepackage{amsmath}

\firstpage{1} 
\pubvolume{1}
\issuenum{1}
\articlenumber{0}
\pubyear{2022}
\copyrightyear{2022}
\datereceived{Nov. 7th, 2022} 
\dateaccepted{TBD} 
\datepublished{TBD} 
\hreflink{https://doi.org/} 

\Title{The ngEHT's Role in Measuring Supermassive Black Hole Spins}

\TitleCitation{The ngEHT's Role in Measuring Supermassive Black Hole Spins}

\Author{Angelo Ricarte $^{1,2}$*\orcidA{}, Paul Tiede$^{1,2}$\orcidB, Razieh Emami$^{1}$\orcidC, Aditya Tamar $^{5}$\orcidD{}, Priyamvada Natarajan $^{2,3,4}$\orcidE{}}

\AuthorNames{Angelo Ricarte, Paul Tiede, Razieh Emami, Aditya Tamar \& Priyamvada Natarajan}

\AuthorCitation{Ricarte, A.; Tiede, P; Emami, R; Tamar, A; Natarajan, P}

\address{%
$^{1}$ \quad Center for Astrophysics | Harvard \& Smithsonian, 60 Garden Street, Cambridge, MA 02138, USA \\
$^{2}$ \quad Black Hole Initiative, 20 Garden Street, Cambridge, MA 02138, USA \\
$^{3}$ \quad Department of Astronomy, Yale University, New Haven, CT 06511, USA\\
$^{4}$ \quad Department of Physics, Yale University, New Haven, CT 06520, USA\\
$^{5}$ \quad Independent Researcher, Delhi-110092, India
}

\corres{Correspondence: angelo.ricarte@cfa.harvard.edu}

\abstract{While supermassive black hole masses have been cataloged across cosmic time, only a few dozen of them have robust spin measurements. By extending and improving the existing Event Horizon Telescope (EHT) array, the next-generation Event Horizon Telescope (ngEHT) will enable multifrequency, polarimetric movies on event horizon scales, which will place new constraints on the space-time and accretion flow.  By combining this information, it is anticipated that the ngEHT may be able to measure tens of supermassive black hole masses and spins.  In this white paper, we discuss existing spin measurements and many proposed techniques with which the ngEHT could potentially measure spins of target supermassive black holes.  Spins measured by the ngEHT would represent a completely new sample of sources that, unlike pre-existing samples, would not be biased towards objects with high accretion rates.  Such a sample would provide new insights into the accretion, feedback, and cosmic assembly of supermassive black holes.}


\begin{document}

\section{Introduction}
\label{sec:intro}

Astrophysical supermassive black holes (SMBHs) can be fully described by just two parameters, their mass (which we denote as $M_\bullet$) and their dimensionless spin parameter (which we denote as $a_\bullet$) \citep{Kerr1963}.  A variety of techniques ranging from dynamical modeling to calibrated scaling relations to broad emission lines have been developed to estimate SMBH masses across the Universe \citep[e.g.,][]{Peterson1993,Wandel+1999,Greene+2010}, as far out to redshifts of $z\sim 6-7$ \citep[e.g.,][]{Fan+2003,Banados+2018,Wang+2021}. These investigations reveal that SMBH masses correlate well with several properties of their host galaxies, most famously leading to tight empirical relationships between SMBH mass and bulge mass, as well as velocity dispersion \citep{Magorrian+1998,Ferrarese&Merritt2000,Gebhardt+2000,Tremaine+2002,Gultekin+2009}.  This suggests growth of SMBHs and their hosts occurs in tandem, as gas is transported to galactic nuclei to form stars and grow SMBHs, and SMBHs inject energy into their hosts in the form of radiation, winds, and jets \citep{Kormendy&Ho2013,Heckman&Best2014}.  Indeed, virtually all current models of cosmic galaxy evolution include SMBH growth and active galactic nucleus (AGN) feedback as a necessary ingredient for suppressing star formation in the most massive host galaxies to a level consistent with observations \citep[e.g.,][]{Vogelsberger+2013,Schaye+2015,Khandai+2015,Dubois+2016,Steinborn+2016,Tremmel+2017,Dave+2019,Ni+2022}.

Compared to their masses, much less is known observationally about SMBH spins. For actively accreting SMBHs with geometrically thin and optically thick accretion disks (those with Eddington ratios roughly in the range $0.01 \lesssim f_\mathrm{edd} \lesssim 0.3$), spin can be estimated from their spectral properties.  A black hole's innermost stable circular orbit shrinks as a function of spin, which leads to higher temperatures and stronger Doppler effects, the latter of which is seen most clearly in the shape of the Iron K-alpha line in X-ray spectra \citep{Remillard&McClintock2006,Brenneman2013}.  Dozens of SMBH spins have been measured using the X-ray reflection spectroscopy technique, which involves modeling X-ray spectra by convolving a rest-frame spectrum with relativistic broadening and redshift effects.  These investigations have found that most SMBHs to which this technique has been applied are highly spinning \citep{Reynolds2021}, with hints of decreases at both high and low masses \citep{Reynolds2021,Mallick+2022}.  However, since accretion directly affects a SMBH's spin, these high spin measurements may be biased and not representative of the spin distribution of the overall SMBH population.  

Little is known observationally about the spins of more typical and ubiquitous, low Eddington rate black holes.  In the case of our own galaxy, \citet{Fragione&Loeb2020} argue that the co-existence of two stellar disks at the galactic center places an upper limit on the spin of Sagittarius A* of $a_\bullet \lesssim 0.1$ needed to prevent Lense-Thirring precession from disrupting them. EHT observations indirectly rule out certain spin values via near-horizon mapping of the accretion flow \citep{EHTC+2019e,EHTC+2021b,EHTC+2022e}, but more theoretical work is ongoing to better understand how reliably spin maps onto horizon-scale observables in the presence of significant model uncertainties.

Quasi-periodic Oscillations (QPOs) are sometimes detected when analyzing light curves of AGN or stellar mass black holes, and their characteristic timescales can also be related to the orbital timescale of matter orbiting in the innermost region of the accretion disc \cite{Dokuchaev:2014,Brink:2015,Dolence:2012}. However, the origin of QPO's is poorly understood, and spin measurements originating from QPOs rely on uncertain origin models that range from disk oscillations \cite{Kato:2010} to the orbital motion of hotspots (\citep[e.g.,][]{Dovciak, BroderickLoeb2005, Broderick2006, Eckart2006, Dokuchaev:2014}). For Very Long Baseline Interferometry (VLBI) observations, tracking the motion of hotspots produced during flares \citep{Gravity+2020} is a promising approach. Closure quantities alone have been shown to be sensitive to periodicities in orbiting hotspots \citep{Doeleman:2009}. By strategically placing additional antennas around the globe to produce denser $uv$ coverage, the next-generation Event Horizon Telescope (ngEHT) will be able to resolve orbital motion. ngEHT will probe the largest SMBHs on the sky with unprecedented spatial and temporal resolution.  By directly observing regions where general relativistic effects are strong, it may recover exquisite constraints on the space-time of these SMBHs and their properties, including their spins.  It is anticipated that the Phase I ngEHT array will enable dozens of mass and spin measurements (Pesce et al.~in prep.). 

In this white paper, we discuss several proposed methods by which the ngEHT could provide novel measures of the spin, and then discuss our current understanding of SMBH spin in the context of the cosmic co-evolution of SMBHs and their host galaxies. Since these new methodologies are completely independent of existing techniques, and because they would be subject to very different selection effects, we argue therefore that even a handful of spin measurements would be greatly impactful for understanding both SMBH accretion flows and their cosmic co-evolution with their host galaxies. In particular, the nearby ngEHT accessible source SMBHs are expected to be accreting preferentially at lower rates than AGN and therefore these measurements will provide a new window into the overall spin distribution of the more characteristic SMBHs. 

\section{Novel Techniques to Infer Spin with ngEHT}
\label{sec:spin_ngeht}

Inferring spin from spatially resolved polarimetric observables and movies of EHT/ngEHT sources is the goal of several recent and ongoing investigations.  As discussed above, any spin constraints derived by ngEHT would probe an entirely new sample of objects, a population of more typical low-Eddington ratio sources.  The most direct probes of spin are based on ``sub-images'' of the accretion flow within the photon ring, which are determined directly by the space-time geometry.  More indirect but more easily achievable approaches involve properties of the accretion flow itself, which is affected not only by the space-time geometry, but also magneto-hydrodynamic (MHD) forces.  

\subsection{Spin from Sub-images:  Theoretically Cleaner, Observationally Harder}

\begin{figure*}
  \centering
  \includegraphics[width=\textwidth]{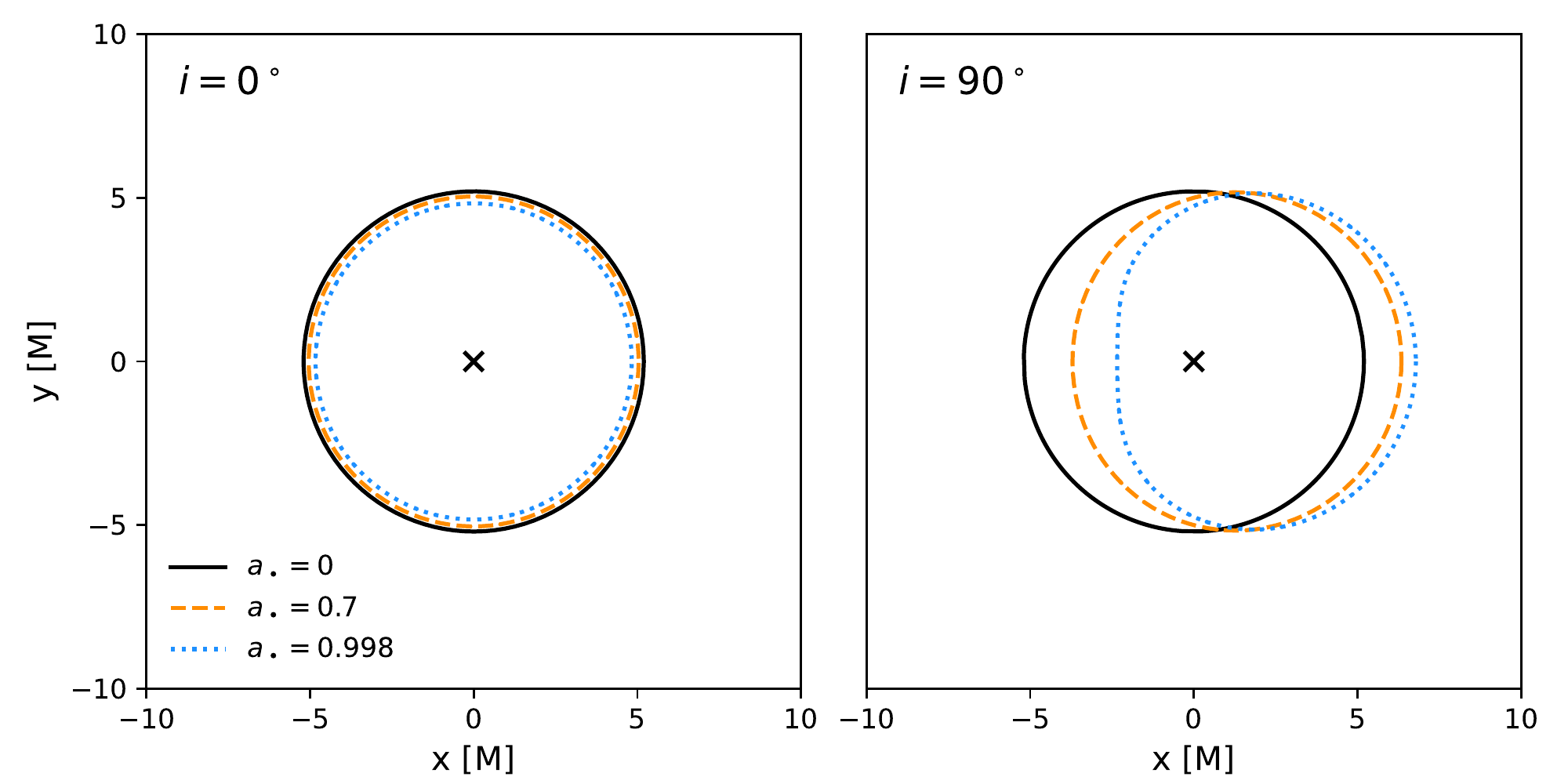}
  \caption{Shape of the $n=\infty$ photon ring or ``critical curve'' as a function of spin and inclination, using the analytic formulae provided in \citet{Chael+2021}.  For face-on viewing angles (left), the critical curve remains circular and shrinks only by about 7\% between 0 and maximal spin.  For edge-on viewing angles (right), the critical curve becomes horizontally displaced and asymmetric as spin increases.}  \label{fig:photon_ring}
\end{figure*}

The most direct impact of spin on black hole images is its effect on the trajectories of photon geodesics, particularly the properties of ``sub-images'' within the photon ring.  A sub-image of a given order is assigned an integer $n$, referring to the number of half-orbits a photon makes around the SMBH on the way to the observer.  The direct image is denoted as $n=0$, while all sub-images with $n \geq 1$ produce  ``photon rings'' in the SMBH image.  Each sub-image is $\approx 4-13 \%$ the width and the flux of the sub-image preceding it, depending on the spin, viewing angle, and position on the ring \citep{Johnson+2020}.

As $n \to \infty$, the shape of the sub-image approaches a ``critical curve'' that is independent of the direct $n=0$ image and completely determined by the space-time \citep{Gralla+2019,Johnson+2020}.  In \autoref{fig:photon_ring}, we plot the shape of the critical curve for SMBHs of different spins for both a face-on and an edge-on viewing angle in geometrized units of $M = GM_\bullet/c^2D$), where $G$ is the gravitational constant, $M_\bullet$ is the SMBH mass, and $c$ is the speed of light, and $D$ is the distance.  For $a_\bullet=0$, this is simply a circle with radius $\sqrt{27} \; M$ since the Schwarzschild metric is spherically symmetric.  For $a_\bullet > 0$ SMBHs, the effect of spin is slight for pole-on viewing angles, decreasing the radius of the critical curve by only $\approx 6\%$.  The effect of spin is more noticeable for an edge-on viewing angle, where the critical curve shifts and grows more asymmetric as a function of spin.  

Directly resolving the width of even the $n=1$ photon ring already requires much finer spatial resolution than accessible to a millimeter array restricted to the size of the Earth, motivating Very Long Baseline Interferometry (VLBI) experiments in space \citep{Johnson+2020}.  Unfortunately, we view M87$^*$ \citep{Walker+2018} and seemingly also Sgr A* at pole-on viewing angles \citep{Gravity+2020,EHTC+2022e} for which the spin signature requires the most precise imaging measurements.  Finally, although the shape of the $n=\infty$ sub-image is fully determined by the space-time, low-order photon rings exhibit a non-negligible dependence on the emission geometry.  It is still possible to break degeneracies and place constraints on spin (as well as mass) with precision measurements of the diameters of the first few sub-images \citep{Broderick+2022a}.  With high spatial resolution and dynamic range, a measurement of the ``inner shadow,'' the lensed image of the equatorial horizon, can also be used to break degeneracies \citep{Chael+2021}.  In the near future, these methods may rely on ``super-resolution'' techniques, imposing strong constraints on the image in an approach between direct imaging and modeling to outperform the nominal spatial resolution of an array \citep{Broderick+2020,Broderick+2022b}.

\subsection{Spin from Accretion Flows:  Theoretically Dirtier, Observationally Easier}

Although sub-images offer the cleanest constraints on the space-time, the signal is weak for face-on viewing angles, and sub-images are extremely narrow and faint.  Alternative approaches for inferring spin are emerging that may utilize properties of the plasma embedded in the space-time.  Inferring spin from the plasma structure and dynamics is less clean than through photon trajectories, since (i) plasma is affected by MHD forces and therefore is not restricted to flow along geodesics and (ii) the emitting region does not necessarily trace the bulk dynamics of the plasma.  Nevertheless, several recent works suggest that this may be a promising avenue, at least in magnetically arrested disk (MAD) systems, requiring much lower spatial resolution than would be necessary to resolve the photon ring.  Through frame-dragging, spin can govern the average magnetic field structure, the morphology of infalling streams, and the orbital motion and appearance of hotspots.
 
\begin{figure*}
  \centering
  \includegraphics[trim=0 80 0 100, clip, width=\textwidth]{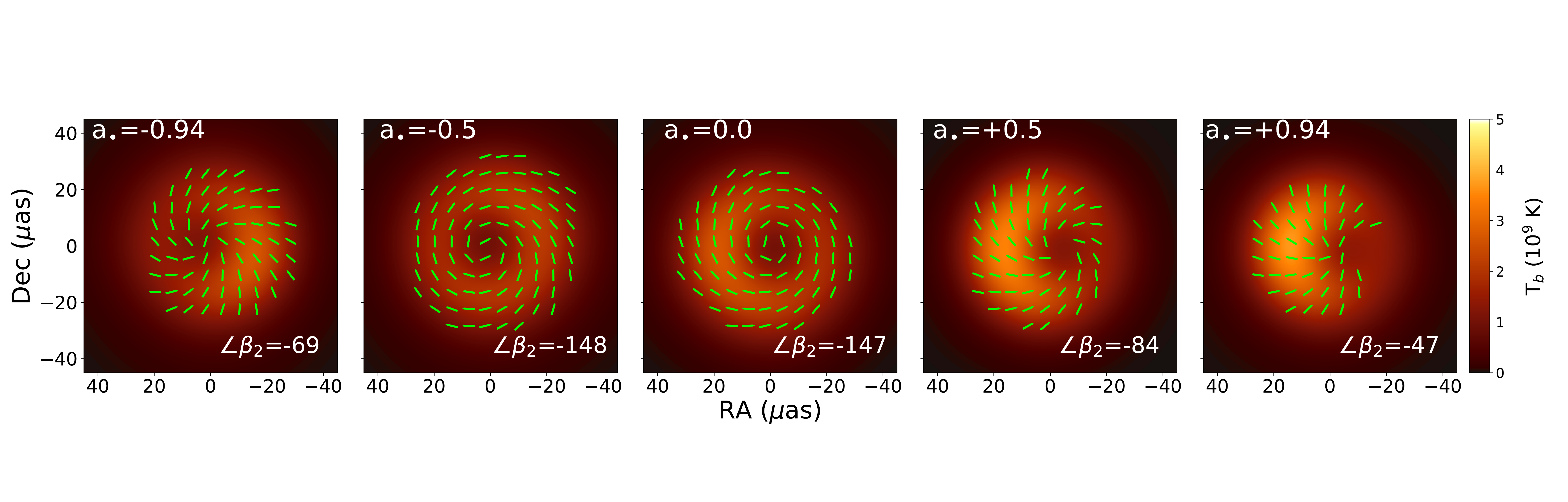}
  \caption{Time-averaged polarimetric images of M87* GRMHD models with increasing spin from left to right, $a_\bullet \in \{-0.94,-0.5,0,0.5,0.94\}$ reproduced from \citet{Emami+2022}, using models presented in \citet{EHTC+2021b}.  The morphology of these linear polarization ticks originate from an evolving magnetic field structure as a function of spin.  Models with higher spin have more toroidal magnetic fields due to frame dragging, which leads to a more radial polarization pattern, while the opposite is true for $a_\bullet \sim 0$.  This is reflected in $\angle \beta_2$, written in degrees at the bottom of each panel, which is closer to $-180^\circ$ for toroidal patterns like the $a_\bullet=0$ model, but moves towards $0^\circ$ for more radial patterns like the $a_\bullet=0.94$ model \citep{Palumbo+2020}}  \label{fig:twistiness}
\end{figure*}

General relativistic magnetohydrodynamic (GRMHD) simulations, wherein magnetized plasma is allowed to evolve in a Kerr space-time, are key numerical tools used to interpret EHT data by evolving plasma and integrating polarized radiative transfer self-consistently \citep{EHTC+2019e,EHTC+2022e}.  \citet{Palumbo+2020} studied the morphology of linearly polarized images of GRMHD models of M87* and found that the twisty morphology of these ticks, quantified by a parameter $\beta_2$, could be used to discriminate between strongly and weakly magnetized accretion disks.  Moreover, as we illustrate in \autoref{fig:twistiness}, the pitch angle of these ticks demonstrates a clear spin dependence.  Recent detailed work delving into the origin of this signal has found that it originates directly from the changing magnetic field structure of the plasma as a function of spin \citep{Emami+2022}. In this picture, frame dragging pulls along plasma which advects magnetic fields along with it. Larger spins result in a magnetic field in the mid-plane which is more toroidal and wrapped in the direction of the SMBH's spin. Then, since synchrotron emission is linearly polarized perpendicular to the direction of the magnetic field, the magnetic field geometry is imprinted onto linear polarization ticks. Indeed, machine learning algorithms point towards this twisty linear polarization morphology as the most important feature for inferring spin (Qiu et al.~in prep.).  Image asymmetry also emerges a spin indicator, reflecting increased Doppler beaming in systems with higher spin \citep{Medeiros+2022}.  

\begin{figure*}
  \centering
  \includegraphics[width=\textwidth]{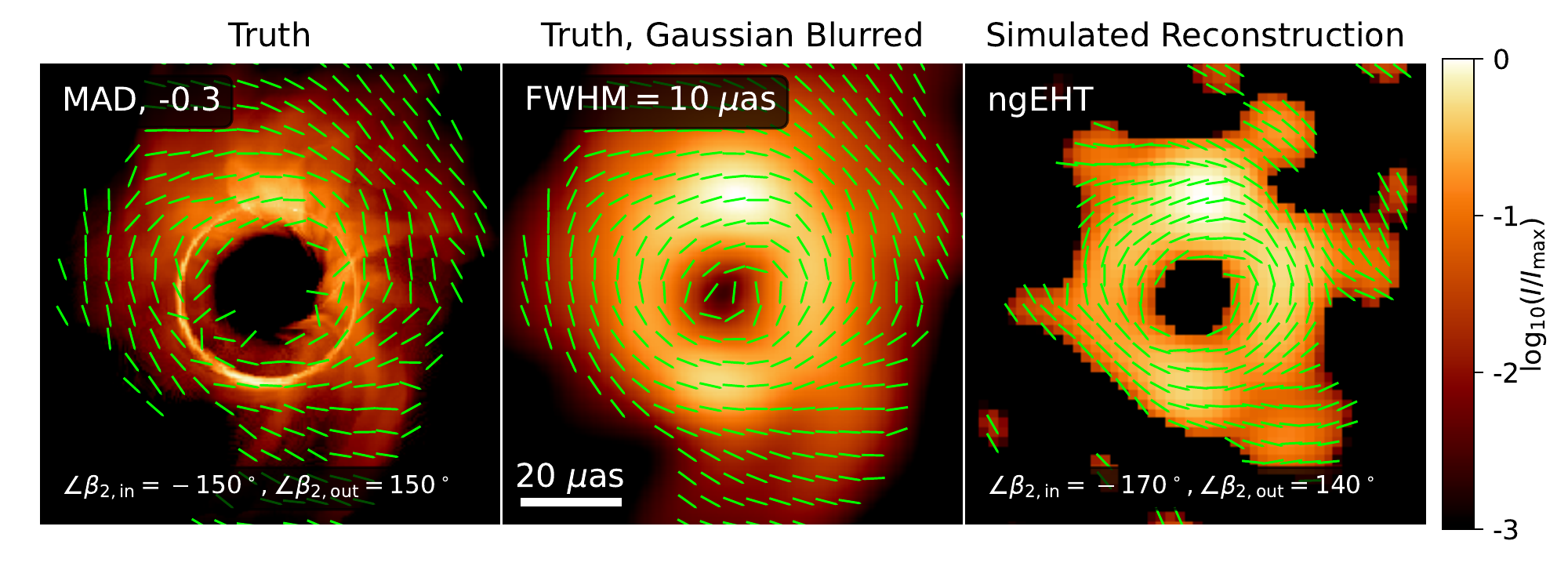}
  \caption{Example total intensity and linear polarization map of a model of M87* that exhibits direct signatures of frame dragging, adapted from \citet{Ricarte+2022}.  Inflowing streams approaching the horizon must turn around due to the existence of an ergosphere in the Kerr space-time, leading to characteristic ``S''-shaped streams and flips in the sign of $\angle \beta_2$ as a function of radius.  The central panel shows the same simulation blurred with a Gaussian with a full-width at half-maximum of 10 micro-arcseconds, which resembles the spatial resolution obtained when simulating the image reconstruction process.  The right-most panel shows a simulated image reconstruction using the phase I ngEHT.  Although the spatial resolution is not sufficient to observe the turnaround of individual streams, the ngEHT can observe the flip in the sign of $\angle \beta_2$ as a function of radius.}
  \label{fig:frame_dragging}
\end{figure*}

It may also be possible to observe frame dragging directly.  In systems where the disk and black hole angular momenta are misaligned, frame dragging can impart a characteristic ``S''-shaped signature onto infalling streams.  Due to magnetic flux freezing in ideal GRMHD, a similar signature can also be imparted onto the linear polarization \citep{Ricarte+2022}.  An example retrograde accretion flow model of M87* is shown in \autoref{fig:frame_dragging}, including a simulated image reconstruction using the Phase I ngEHT array.  The ngEHT could observe the sign flip in $\angle \beta_2$ across the photon ring corresponding to a turnaround in the accretion flow.  At larger radii, misalignment is generally expected between the SMBH angular momentum axis and that of inflowing gas.  As a result, accretion disks can warp, tear, and potentially undergo Lense-Thirring precession \citep[e.g.,][]{Fragile+2007,Liska+2021}.  With sufficient dynamic range in both intensity and spatial scale, this may result in a visible transition at some radius, or potentially impart a QPO signal in the time domain, meriting further study.  Frame dragging may also allow spin measurements if a pulsar is discovered in close proximity to Sgr A* \citep[e.g.,][]{Wex&Kopeikin1999,Pfahl&Loeb2004,Psaltis+2016,DeLaurentis+2018}, although transient searches have not yet uncovered one \citep{Kuo+2021,Mus+2022}.

\begin{figure}
    \centering
    \includegraphics[width=\textwidth]{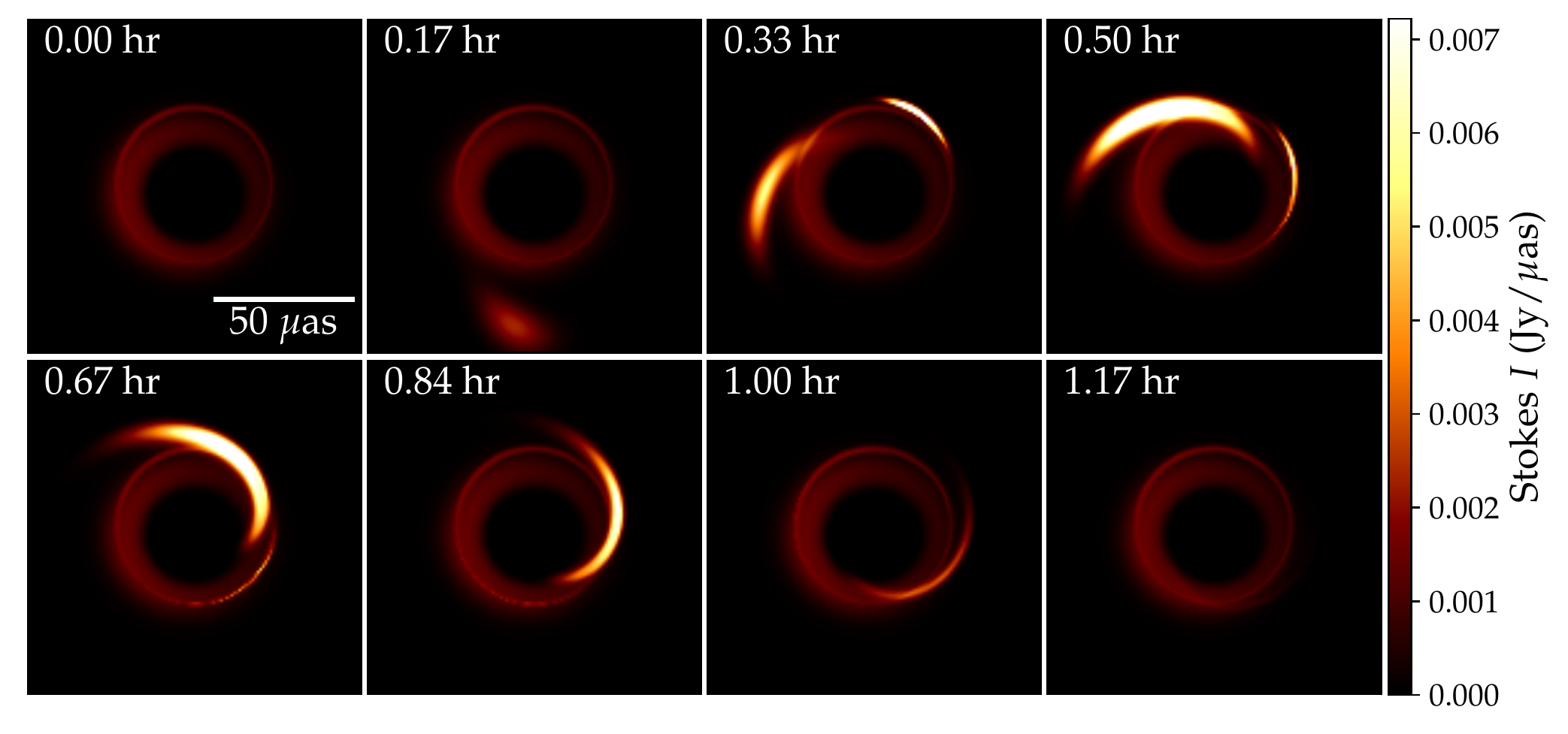}
    \caption{Frames of a hotspot simulation using the model from \citet{Tiede+2020}, where a shearing hotspot is superimposed on top of a static accretion flow. By tracking the motion of the hotspot, the ngEHT could constrain the dynamics of the accretion flow. Additionally, while the motion of the primary hotspot is highly dependent on the accretion flow, the appearance of the secondary image probes the black hole spacetime \cite{Hadar, Wong2021} allowing for a direct measurement of the black holes spin and inclination.}
    \label{fig:hotspot_movie}
\end{figure}

\subsection{Spin from the Time Domain: Movies and Motion}

Lacking spatial resolution, it may also be possible to extract spin with excellent temporal resolution. Due to lensing, light that travels from the black hole to an observer travels along multiple paths. Spatially this leads to multiple sub-images or light echoes. However, these echoes appear around the black hole's photon ring at different times. \citet{Wong2021} demonstrated that measuring the light echoes or glimmer location in time and angle around the black hole precisely encodes the spin and inclination of the central black hole. The utility of temporal measurements was noticed earlier by \citet{BroderickLoeb2005}, who demonstrated that even the fractional polarization light curve during a flare was very sensitive to the inclination of the black holes accretion disk. The autocorrelation structure of glimmer was further analyzed in \citep{Chesler2020, Hadar}. \citet{Hadar} and \citet{Wong2021} demonstrated that measuring the angular location and arrival time of glimmer is feasible for finite-resolution images like those produced from the EHT.  Conceptually, excellent snapshot imaging substitutes for exquisite spatial resolution for measuring glimmer. This means that one must quickly construct SMBH images with high dynamic range (more than an order of magnitude to measure the sub-image) on timescales similar to the light-crossing time of the black hole, which is on the order of minutes for Sgr A$^*$.  So far, studies of SMBH glimmer have been limited to simple toy models, motivating additional study to better understand observational requirements in the presence of realistic stochastic accretion flow.

Finally, movies capturing the dynamics of the plasma in the vicinity of the event horizon made by ngEHT will also allow us to access dynamics more directly in the time domain. \citet{Moriyama,Tiede+2020} demonstrated that by directly modeling the appearance and evolution of hotspots seen by GRAVITY \citep{Gravity+2020} around Sgr A*, the EHT/ngEHT could measure both the black hole spin and accretion dynamics. Follow-up work by \citet{levis} demonstrated that the ngEHT could potentially measure an arbitrarily complicated emissivity profile using similar methods, although they assumed a fixed spacetime and hotspot velocity field. However, these direct modeling methods are still quite restrictive in terms of types of magnetic fields and velocity fields used to describe the hotspot. Additionally, both \citet{Tiede+2020} and \citet{levis} ignored the surrounding stochastic accretion flow. More research into the expected velocity field of hotspots and how the surrounding accretion flow impacts measurements of plasma dynamics and black hole spin is needed.

Another avenue to measuring plasma dynamics is to extract the motion of the on-sky image instead of around the black hole. The direct image domain approach tends to be computationally simpler, and several different dynamical models \citep{dynamicalimg, arrasM87, starwarps} exist. Additionally, these models make fewer assumptions about the nature of the motion compared to the direct modeling approach described above and are agnostic about the underlying physics. Emami et al.~(in prep.) presents an initial exploration demonstrating the feasibility of tracking hotspots around Sgr A* using the ngEHT. The downside of the image domain approach is that relating the on-sky motion to the dynamics of the plasma and surrounding spacetime is poorly understood and requires additional research. In the end for actual observations, both the more direct but restricted parametric modeling and the flexible but less specific image domain modeling will be necessary to test of the robustness of any measurements to different modeling choices.

At present, inferring spin from accretion flow properties is limited not only by observational limitations, but also in large part by theoretical uncertainties.  A GRMHD-based analysis of the polarized EHT image of M87* already suggestively rules out certain spin values, but uncertainties regarding electron heating and cooling currently limit our conclusions \citep{EHTC+2021b}.  These uncertainties propagate into the geometry of the emitting region and Faraday rotation, both of which are integral for interpreting polarized data.  Theoretical developments in this area could significantly improve spin constraints by reducing the allowable parameter space.

\section{Implications of SMBH Spin}
\label{sec:implications}

A SMBH grows via both accretion (which may include multiple triggering mechanisms and accretion modes) and mergers with other SMBHs. Its spin encodes recent gas dynamical activity determined by its mode of accretion as well as its merger history. Accretion via a thin disk for instance, imparts angular momentum in the direction of the disk's angular momentum. Retrograde accretion therefore spins a SMBH down, while prograde accretion for a SMBH surrounded by a thin disk can spin up a SMBH up to the theoretical maximum of $a_\bullet=0.998$ \citep{Thorne1974}. For geometrically thick disks, on the other hand, energy extracted to power jets via the \citet{Blandford&Znajek1977} process can cause spin down even in the prograde case, which may have interesting implications for SMBHs imparting ``maintenance-mode'' feedback for Gyrs \citep{Narayan+2022}. SMBHs may also accrete chaotically, for example from the stocastic scattering of molecular clouds with random angular momenta, which would decrease spin on average \citep[e.g.,][]{King+2008,Volonteri+2013}. Finally, SMBH-SMBH mergers also impact the remnant's spin, depending on the mass ratio and the relative alignments between the orbital angular momentum vector and the spin vectors of the two SMBHs \citep[][]{Rezzolla+2008}. An equal mass merger of SMBHs without pre-existing spins will tend to produce a remnant with a spin of $a_\bullet \sim 0.7$, but as with accretion, many low-mass mergers on random orbits will decrease the spin \citep{Volonteri+2005,Berti&Volonteri2008}. All of these complexities can now be modeled self-consistently in semi-analytic models, and dramatically different results can be obtained depending upon one's assumptions about the alignment of accretion disks and mergers. 

As a demonstration, we compute spin probability distributions for SMBHs hosted in 100 different $10^{15} \ M_\odot$ halos (like M87*) using the simple semi-analytic model for SMBH evolution developed in \citet{Ricarte&Natarajan2018a,Ricarte&Natarajan2018b,Ricarte+2019}.  We build upon the \citet{Ricarte+2019} model by including spin evolution by accretion and mergers self-consistently with mass assembly as in previous works \citep[e.g.,][]{Volonteri+2005,Berti&Volonteri2008}.  In this model, accretion is triggered by halo mergers with a mass ratio of 1:10 or larger.  When this occurs, an Eddington ratio is drawn from a distribution appropriate for Sloan Digital Sky Survey (SDSS) broad line quasars \citep{Kelly&Shen2013}, near Eddington including a super-Eddington tail.  With minimal assumptions, this model reproduces the bolometric luminosity function of AGN out to $z=6$ quite well \citep{Ricarte+2019}.  Here, spin is evolved using analytic calculations appropriate for a thin disk \citep{Bardeen1970} up to a maximum value of $a_\bullet=0.998$ \citep{Thorne1974}.  For this simple, illustrative calculation, we assume that merger-triggered accretion always occurs via prograde thin disks.  Following a SMBH-SMBH merger, which we assume is randomly aligned, we use the equations of \citet{Rezzolla+2008} to compute the spin of the remnant.

We isolate one of the theoretical uncertainties that affects the cosmic evolution of SMBH spin: the probability that a SMBH merger occurs following a halo merger.  The most massive SMBHs in the universe are especially sensitive to this astrophysics, as both cosmological simulations and semi-analytic models predict that SMBH-SMBH mergers can in fact dominate the final mass budget of these SMBHs \citep{Ricarte&Natarajan2018a,Weinberger+2018,Pacucci&Loeb2020}.  However, the journey between halo/galaxy merger and SMBH merger involves traversing many orders of magnitude in spatial scale, and requires multiple physical mechanisms from dynamical friction on the largest scales to gravitational wave emission on the smallest scales \citep{Begelman+1980,Colpi2014}.  When a major galaxy merger occurs, the central SMBHs may not merge for a variety of reasons, including kilo-parsec scale wandering owing to a messy and cosmologically evolving potential \citep[e.g.,][]{Tremmel+2017,Bortolas+2020,Izquierdo-Villalba+2020,Ricarte+2021a}, potential delays around one parsec when neither dynamical friction nor gravitational wave emission are efficient \citep{Milosavljevic&Merritt2001}, or even multi-body scatterings in the galactic nucleus \citep{Volonteri+2003}.  To simply and clearly illustrate our model's sensitivity to SMBH-SMBH mergers, we vary a constant SMBH merging probability following a halo merger, for which we select 3 values, $p_\mathrm{merge} \in \{0.1,0.3,1.0\}$.  We keep this probability equal to 0 if the halo merger had a mass ratio more extreme than 1:10, in which case the satellite should be stripped, leaving the central SMBH in the outskirts of the halo.

\begin{figure*}
   \centering
   \includegraphics[width=\textwidth]{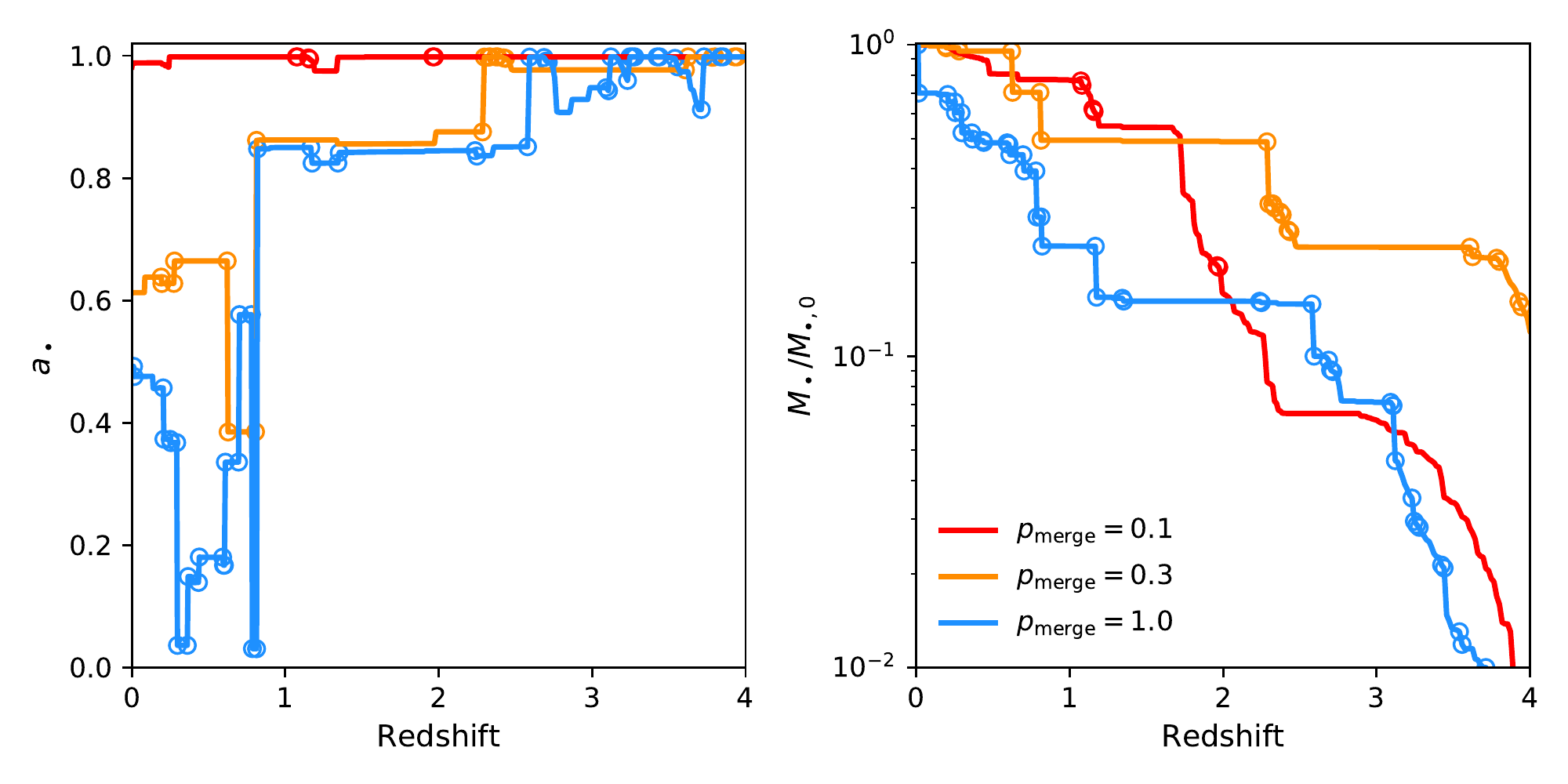} 
   \caption{Example evolutionary histories of SMBH spin (left) and mass normalized by final mass (right) as a function of redshift in $10^{15} \ M_\odot$ halos using a semi-analytic model for cosmological SMBH assembly.  We demonstrate sensitivity to the merger history of SMBHs by varying a free parameter $p_\mathrm{merge}$, which sets the probability that a SMBH merger occurs following a major halo merger.  Different colors encode different values of $p_\mathrm{merge}$, and open circles mark SMBH-SMBH mergers with mass ratios of at least 1:100.  Many mergers occur in these massive halos, which can cause sharp jumps in $a_\bullet$ if $p_\mathrm{merge}$ is large.  If $p_\mathrm{merge}$ is small, $a_\bullet$ stays near the maximum value of $a_\bullet=0.998$.
   \label{fig:sam_tracks}}
\end{figure*}

In \autoref{fig:sam_tracks}, we plot three representative example SMBH assembly histories as a function of redshift for our different values of $p_\mathrm{merge}$.  In the left panel, we plot the evolution of $a_\bullet$, and in the right panel, we plot the evolution of $M_\bullet/M_{\bullet,0}$, where $M_{\bullet,0}$ is the final SMBH mass at $z=0$.  Open circles mark SMBH mergers with a mass ratio of at least 1:100.  Many may occur at low-redshift for these massive halos if $p_\mathrm{merge}$ is large, which can lead to sharp jumps in $a_\bullet$.  On the other hand, if $p_\mathrm{merge}$ is small, $a_\bullet$ stays near its maximum value of $a_\bullet=0.998$, since all accretion is assumed to occur via prograde thin disks.  The SMBH in the model with $p_\mathrm{merge}=1$ assembles its final mass latest in cosmic time, as many SMBH mergers contribute to its final mass budget. 

\begin{figure*}
   \centering
   \includegraphics[width=\textwidth]{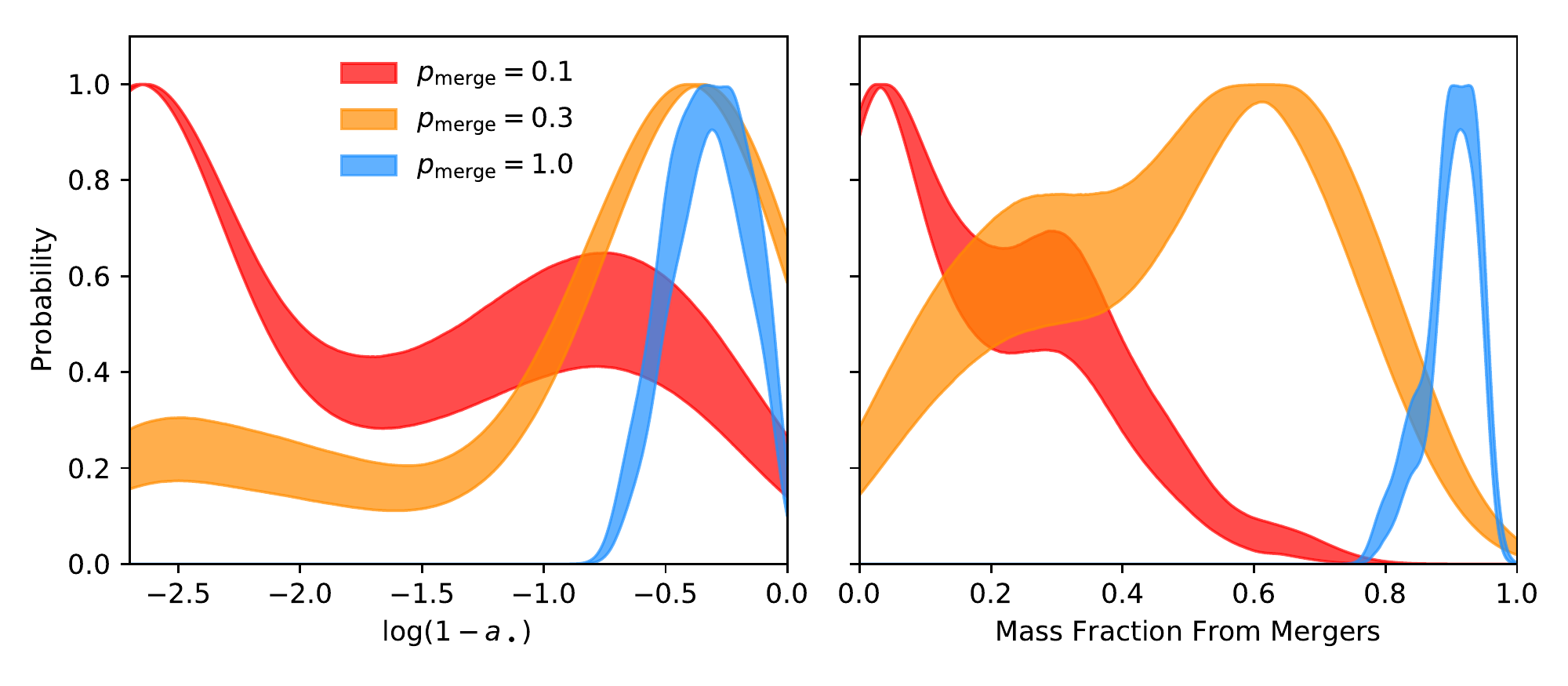} 
   \caption{Distributions of spin (cast in terms of $\log (1-a_\bullet)$) and mass fraction accumulated via mergers for SMBHs in $10^{15} \ M_\odot$ halos using a semi-analytic model.  In the left panel, we see that models which allow more mergers produce SMBHs with less extreme spin values.  In the right panel, we see that most SMBH mass can originate from mergers if all of them are allowed to occur.
   \label{fig:sam_distributions}}
\end{figure*}

Using all 100 different assembly histories that we have computed, we then plot distributions of spin (left; now plotted in terms of $\log (1-a_\bullet)$) and the fraction of the final mass accumulated via mergers (right) in \autoref{fig:sam_distributions}.  As expected, these three different values of $p_\mathrm{merge}$ yield different spin distributions, with more maximal spins in the model with the fewest mergers.  In the right panel, we see that for models with $p_\mathrm{merge}=1$, over 80\% of the final mass is accumulated via SMBH-SMBH mergers.  In this model, this is because their gas-driven accretion is occurs very early in the universe in order to produce a sufficient quantity of luminous quasars at $z=6$.  Then, the model then shuts off gas-driven growth at later times, so as not to overshoot the $M_\bullet-\sigma$ relation, but not SMBH-SMBH mergers.  

Apart from cosmic evolution, a SMBH's spin also has an immediate impact on its accretion and feedback processes. The radiative efficiency of a thin disk is strongly sensitive to the SMBH's spin, reaching up to $\epsilon=42\%$ for $a_\bullet=1$ compared to a mere $\epsilon=6\%$ for $a_\bullet=0$. For geometrically thick disks, jet efficiencies also scale strongly with spin, and may even exceed 100\% for prograde disks approaching $a_\bullet=1$ that power jets via the Blandford-Znajek mechanism \citep{Tchekhovskoy+2011,Narayan+2022}. A sample of SMBHs with both spin and jet power measurements by the ngEHT could help elucidate the mechanism that powers jets.  Finally, following a SMBH-SMBH merger, spin and orbital energy can be converted into a velocity kick, which may offset SMBHs from their host's centers. These kick velocities have been computed from general relativistic simulations and are found to range between 100-1000 km s$^{-1}$. Therefore, in some cases where the kick velocity exceeds the velocity dispersion of the galaxy's gravitational potential, the remnant can be ejected from the nucleus \citep[e.g.,][]{Volonteri&Perna2005}.

\section{Summary}
\label{sec:summary}

Robust SMBH spin measurements have only been inferred for about twenty sources, and this sample is methodologically biased towards objects with high accretion rates.  With spatially resolved polarimetry and time-domain information, the ngEHT has the potential to place spin constraints on a completely different sample of SMBHs, with very different model assumptions and selection effects.  Spins measured by ngEHT would sample an ideal type of source: typical low-Eddington ratio objects with large masses, responsible for maintenance mode AGN feedback and whose growth history can be majorly impacted by SMBH-SMBH mergers.  Spin constraints of such a sample would have implications for the accretion, feedback, and cosmic assembly of SMBHs.

The cleanest probes of spin for ngEHT sources involve ``sub-images'' of accretion flows in what is referred to as the photon ring of optically thin SMBH images.  These methods are clean because the paths that photons take is determined entirely by the space-time.  The signature of spin in the image of the photon ring is quite subtle except for the edge-on cases with substantial spin.  With time domain information, we may also constrain spin through characteristic light echoes or ``glimmer.''  More indirect but more observationally accessible probes of spin rely on the structure and motion of the plasma in the accretion flow, which are affected by the forces of MHD in addition to gravity.  We have discussed how the frame-dragging of plasma threaded with magnetic fields as a function of spin can impart a signature in linear polarization ticks.  Inflowing streams may also exhibit changes in pitch angle, even flipping handedness with radius, depending on the spin.  In the time domain, the apparent motion of hotspots can also encode the underlying geometry.  These indirect and model-dependent spin inferences would benefit from continued theoretical development in the upcoming years to determine sensitivity to initial conditions and model assumptions.

The prospect of measuring spin motivates event horizon scale polarimetry with high spatial resolution, high dynamic range, and fine temporal sampling.  For each of these, the $uv$ coverage provided by the ngEHT will be essential.

\section{Acknowledgments}

We thank the National Science Foundation (AST-1716536, AST-1935980, AST-2034306, AST-1816420, and OISE-1743747) for financial support of this work.  This work was supported in large part by the Black Hole Initiative, which is supported by grants from the Gordon and Betty Moore Foundation and the John Templeton Foundation. The opinions expressed in this publication are those of the author(s) and do not necessarily reflect the views of the Moore or Templeton Foundations. RE acknowledges the support from NASA via grant HST-GO-16173.001-A.

\bibliography{ms.bib}

\end{document}